\documentstyle[amssymb,epsf,wrapft,twoside,mbf]{ptptex}
\notypesetlogo  
\title{\hfill{{\small{FZJ-IKP(TH)-2000-19}}}\\
\vspace{0.8cm}
Scalar Mesons and Chiral Symmetry
}
\author{%
J. A. {\sc{Oller}}}
\inst{%
 Forschungszentrum J\"ulich, Institut f\"ur Kernphysik (Theorie), D--52425 J\"ulich, Germany}
\abst{%
It is the purpose of the present manuscript to emphasize those aspects that make the 
scalar 
sector with vacuum quantum numbers rather unique. Chiral symmetry is the basic tool for 
our study together with a resummation of Chiral Perturbation Theory (CHPT) 
that stresses the role of unitarity but also allows one to include explicit resonance 
fields and to match with the CHPT expansion at low energies.
\begin{center}
\vskip 15pt
{\scriptsize {\it Invited Talk at the $\sigma$--meson workshop, Kyoto, June 2000}}
\end{center}
}

\begin{document}

\maketitle

\setcounter{tocdepth}{4}

\section{Introduction}

In the $J^{PC}=0^{++}$ sector hadrons really interact strongly. In fact, one 
can find in the 
literature  a consensus with respect to the strong unitarity corrections 
(rescattering effects) that take place in this sector 
\cite{weis,au,gassulf,ulf,torn,juelich,oo99}. 
As a 
matter of fact, this strong rescattering usually involves different channels with very 
pronounced cusp structures at the opening of the heavier thresholds coinciding with 
the presence of resonance states: For the $I=0$ case one has the $f_0(980)$ and 
the $K\bar{K}$ channel, for the $I=1$ the $a_0(980)$ and the $K\bar{K}$ threshold and 
finally in the $I=1/2$ one has the $K\eta'$ channel and the $K^*_0(1430)$ resonance. 
Of course, in these cases simple 
Breit-Wigner forms for resonances are 
not justified and even the appearance of dynamical resonances originated by the 
residual but, in this case, strong meson--meson interactions can take place 
\cite{weis,juelich,oo99,oo97}. In any case, 
these strong unitarity corrections tend to mask the resonance properties by making 
some of them very wide and also by providing large backgrounds that have to be pinned 
down rather precisely. 

In connection with the previous discussions, the appearance of 
large violations of the Okubo--Zweig--Izuka (OZI) rule in the $0^{++}$ sector can 
be expected. The only general and well founded explanation of the OZI rule is 
large $N_c$ QCD \cite{oft} since OZI violating graphs imply the presence of an extra 
quark loop 
diagram and these contributions are suppressed by a factor $1/N_c$. In the large $N_c$ 
limit mesons do not interact 
with each other \cite{oft}, establishing in this way the basis for a weakly 
coupling theory of 
the strong interactions in terms of hadronic degrees of freedom in which the leading 
contributions come from tree-level diagrams (local and pole terms). This picture is, to 
a large extend, highly successful in the vector and tensor channels. However, in the 
$0^{++}$ sector loops, which are responsible for the unitarity of the $S$-matrix, 
are strongly enhanced and then it is 
natural to expect a departure from the large $N_c$ scenario. For instance, 
it is clear that the $0^{++}$ spectrum is not dominated by ideally mixed $\bar{q}q$ 
nonets,
 being the latter a consequence both 
of large $N_c$ 
and the OZI rule.  Notice that the 
OZI rule is one of the key ingredients in simple quark models and has been usually 
advocated as a way to pin down values for low energy constants in Chiral 
Perturbation Theory\cite{leut}.

\section{Crossed channel dynamics and lowest order CHPT}

In this section we want to point out the facts which drive 
the phenomenological behavior of the $I=0,$ 1 and $1/2$ $0^{++}$ channels. 
The first one  has its roots in a violation of the expected large 
$N_c$ results and the second is just a requirement of chiral symmetry.

\subsection{Crossed channel dynamics}

In ref. \cite{oo99} the unphysical cut contributions, due to crossed channel dynamics,  
were estimated making use of the results of ref. \cite{ulfres} in which the 
next-to-leading order CHPT amplitudes were supplied with the exchange of explicit 
resonance fields 
from the chiral symmetric Lagrangians of ref. \cite{pich}.
 In this way, the range of applicability of chiral constraints was enlarged up to around 
0.8 GeV. From the diagrams considered in ref. \cite{ulfres}, one isolates 
those 
corresponding to loops and to the exchange of resonances both in the crossed $t-$ and 
$u-$ channels, figs.1a,b respectively. For further details we refer to section V of ref. 
\cite{oo99}. The important 
result was that up to $0.8$ GeV this set of diagrams amounts to just of a few percent of
 the sum of the $s-$channel exchange of resonances plus the 
lowest order CHPT amplitude, figs.1c,d respectively. This result has been recently 
corroborated in ref. \cite{jamin}.

As discussed in ref. \cite{oo99} the smallness of the unphysical cut contributions in 
the physical region is 
a consequence of a large cancelation between the crossed loops and the crossed 
exchange of resonances, otherwise these contributions would be important.  That this 
cancelation takes place is a clear signal of 
large $N_c$ violation in these channels because loop physics is suppressed by an 
extra $1/N_c$ 
factor with respect to the tree-level exchange of vector plus scalar resonances.

\subsection{Lowest order CHPT}

In ref. \cite{oo99} the general structure of a partial wave amplitude when the 
unphysical cuts are discarded was shown to be:

\begin{equation}
\label{nolhc}
T=\left[\sum_{i} \frac{\gamma_i}{s-s_i}+a^L+g(s) \right]^{-1}
\end{equation}
where the sum extends over the CDD poles\cite{casti}, $a^L$ is a leading subtraction 
constant in the large $N_c$ counting as ${\mathcal{O}}(N_c)$, and $g(s)$ is the two 
meson loop function depicted in fig.2 and is ${\mathcal{O}}(N_c^0)$. For the 
case of equal meson masses $m$, $g(s)$ is given by:
\begin{equation}
\label{g}
g(s)=a^{SL}(\mu)+\frac{1}{(4 \pi)^2}\left[\log \frac{m^2}{\mu^2}+ \sigma(s) \log 
\frac{\sigma(s)+1}{\sigma(s)-1}\right]
\end{equation}
 In the 
following we will take $\mu=M_\rho$ with $M_\rho$ the mass of the $\rho$ meson. However, 
it should be clear that our results are scale independent since any change in the 
scale $\mu$ of $g(s)$ is reabsorbed by a change in $a^{SL}(\mu)$.

\begin{figure}[t]
\epsfysize=3.5 cm
\centerline{\epsffile{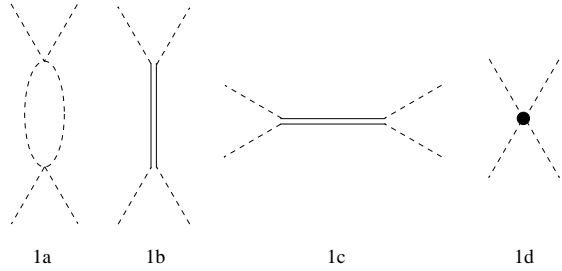}}
\caption{Fig.1a: $t-$ and $u-$channel crossed loops at ${\mathcal{O}}(p^4)$ in CHPT. 
Fig.1b: $t-$ and $u-$channel exchange of scalar and vector 
resonances. Fig.1c: $s-$channel exchange of scalar and 
vector resonances. Fig.1d: Lowest order CHPT.}
\label{fig1}
\end{figure}

Notice that a CDD pole corresponds to a zero of the amplitude $T$ since it is 
a pole in 
the {\it{denominator}}. Interestingly, due to the Goldstone boson nature of the pions, 
the presence of the Adler zeroes follows and hence, at least, one needs a CDD pole per 
partial wave (notice that $a^L$ can be 
considered as a CDD pole at infinity). 
The position of this pole ($s_0$) and its residue ($\gamma_0$) is 
just the information contained in the lowest order CHPT amplitudes. Of course, both 
the position and the residue are given to the corresponding lowest order.

Let us go now to consider the consequences of the presence of the Adler zeroes 
by restricting the sum over the CDD poles in eq. (\ref{nolhc}) to only the one 
that corresponds to the Adler zero for each particular partial wave. We will 
consider below the stability of the results under the inclusion of 
higher CDD poles.

We will also discuss simultaneously the S- and P-wave $\pi\pi$ partial amplitudes 
in order to make manifest the striking differences between both cases.

To reproduce the $\pi\pi$ P-wave scattering data, which is clearly dominated by the 
presence of the $\rho$ resonance, one needs \cite{oo99}:

\begin{eqnarray}
\label{pw}
a^L&\simeq&-\frac{6f_\pi^2}{M_\rho^2}\simeq -9\times 10^{-2} \nonumber \\
a^{SL}&=&0
\end{eqnarray} 
where the analytic expression for $a^L$ follows from Vector Meson Dominance \cite{oo99}.

On the other hand, in order to describe the data for the $\pi\pi$ $I=0$ S-wave below 
0.8 GeV, 
where two pions are still the relevant intermediate states, one has\cite{oo99}:

\begin{eqnarray}
\label{sw}
a^L&=&0 \nonumber \\
a^{SL}&\simeq&- 5\times 10^{-3} 
\end{eqnarray}
From these values it follows as well the presence of the $\sigma$ pole by applying eq. 
(\ref{nolhc}) \cite{oo99,oo97}.

For the $\pi\pi$ P-wave  the Adler zero is just located at threshold while in the 
second case, for the S-wave, is close to $m_\pi^2/2$. The differences between eqs. 
(\ref{pw}) and 
(\ref{sw}) are really astonishing. To further sharpen these statements, notice that 
while one can reabsorb $a^{SL}$ in eq. (\ref{sw}) just by a change of order one in the 
scale 
$\mu$, this is not possible for $a^L$ in eq. (\ref{pw}). In fact, the necessary 
change in the scale $\mu$ in this case is:
\begin{equation}
\label{change}
\mu \rightarrow \exp(- 8 \pi^2 a^L) \mu \simeq 1 \hbox{ TeV}.
\end{equation}
while any `natural' scale $\mu$ should be around 1 GeV.

\begin{figure}[t]
\epsfysize=1.5 cm
\centerline{\epsffile{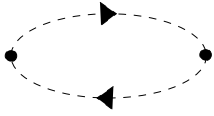}}
\caption{Two meson loop function $g(s)$.}
\label{fig2}
\end{figure}

This result tells us that the $\rho$ is a preexisting state that cannot be generated by 
loop physics, that is, its mass is ${\mathcal{O}}(N_c^0)$. In contrast, the $\sigma$ 
meson generated by applying the results of eq. (\ref{sw}) to eq. (\ref{nolhc}) 
can be considered as a dynamical resonance generated through the iteration of the 
lowest order CHPT amplitude. In this case, its mass counts as 
$f_\pi^2={\mathcal{O}}(N_c)$ and hence, in the large $N_c$ limit, the pole disappears 
by moving to infinity.

Alternatively, we can also present the previous discussion by saying that if we 
wanted to generate a pole with a mass around 0.8 GeV in the $\pi\pi$ P-wave amplitude 
just by the iteration of the lowest order CHPT amplitude ($a^L=0$), we would need a 
scale of around 1 TeV whereas in the S-wave the resulting scale is of the 
order of 1 GeV. This difference arises because of the residues of the CDD poles 
associated to the Adler zeroes. 
While 
in the S-wave one has $f_\pi^2$ in the P-wave the residue is $6f_\pi^2$, that is, 
there is an enhancement of a factor $6$ in the later case with respect to the former. 
Because the scale $\mu$ only appears logarithmically in $g(s)$ all these factors of 
difference imply an exponential scaling in $\mu$. Of course, these residues are 
 just consequence of chiral symmetry.

This interpretation about the dynamical origin of the $\sigma$ 
meson was tested in ref. \cite{oo99} by allowing the presence of explicit resonances, 
or equivalently, incorporating more CDD poles. The result was that after 
fitting the present experimental data, no preexisting resonance appeared in the $I=0$ 
S-wave $\pi\pi$ that could be related to the $\sigma$ pole.

\section{SU(3) and Final State Interactions}

\subsection{SU(3) related channels}

The results discussed in the previous section were generalized to the SU(3) case in 
ref. \cite{oo99} and an improved and very detailed study of the S-wave $K\pi$ scattering 
has been recently given in ref. \cite{jamin}.

In ref. \cite{oo99} the coupled channel version of eq. (\ref{nolhc}) was also derived 
and used to further investigate the whole set of SU(3) connected $I=0,$ 1, 1/2 S-wave 
meson--meson partial amplitudes from threshold up to about 1.3--1.4 GeV. For higher 
energies multiparticles states cannot be further neglected.

In that paper together with the Adler zeroes previously discussed, the inclusion of 
two nonets of preexisting scalar resonances with masses below 1.5 GeV was allowed. 
However, after fitting phase shifts and inelasticities, the couplings of one of these 
scalar nonets were compatible with zero and the fit only required the presence of 
one scalar octet with a mass around 1.4 GeV and of a singlet around 1 GeV. For further 
details see ref. \cite{oo99}. 

The resonance content of the 
solutions was also studied \cite{oo99} and two sets of scalar resonances was observed. 
The octet of preexisting resonances around 1.4 GeV gives rise to eight resonance poles 
with masses very close to the physical resonances $f_0(1500)$, $a_0(1450)$ and 
$K^*_0(1430)$ \cite{pdg}. The singlet resonance around 1 GeV evolves 
continuously from its bare pole position to the final one giving rise to a contribution 
to the  $f_0(980)$ resonance (this can be seen by inserting gradually $g(s)$ in eq. 
(\ref{nolhc}), multiplying it by a factor $\lambda$ which takes values 
between 0, tree-level, and 1, final result).

However, the presence of extra resonance poles that do not originate from 
the set of the preexisting ones was also observed. These comprise the $a_0(980)$ pole 
with $I=1$, the 
$\kappa$ with $I=1/2$ and two poles in the $I=0$ case: the $\sigma$ and a very important 
contribution to the $f_0(980)$ due  to the $K\bar{K}$ threshold. All these poles 
originate just by iterating the lowest order CHPT amplitudes, as discussed in the 
previous section with respect to the $\sigma$ meson. In fact, it was also observed in 
ref. \cite{oo99} that when moving continuously to the SU(3) limit these poles bunch 
together in a degenerate octet plus a singlet.

Ref. \cite{jamin}  was devoted to a thorough study of the S-wave $K\pi$ scattering 
amplitudes. For the $I=1/2$ a description of the data was accomplished up to about 
2 GeV. The input considered in this reference is an improved one with respect to that 
used in ref. \cite{oo99}. The differences are: 1) Unphysical cut contributions were 
included by considering crossed exchange of vector and scalar resonances and crossed 
loops calculated at ${\mathcal{O}}(p^4)$ in CHPT\footnote{In fact, a matching to the 
next-to-leading order $K\pi$ amplitudes is given in ref. \cite{jamin} and at the tree level 
the approach is crossed symmetric.}, 2) the $K\eta'$ channel was included making use 
of a combined chiral and $1/N_c$ expansion \cite{leute}: 
$1/N_c \backsim {\mathcal{O}}(p^2)$. Notice 
that this state is a fundamental one in order to study properly the $K_0^*(1430)$ 
resonance. Despite these improvements the pole of the $\kappa$ resonance remains 
basically in the same position as the one found in ref. \cite{oo99}. Furthermore, 
in this latter reference a perturbative expansion in terms of the unphysical cuts 
contributions 
for the $I$=0, 1 and 1/2 S-wave meson--meson amplitudes  
was argued and the overall agreement between this reference and ref. 
\cite{jamin} gives further support to this point of view. We remind the reader what was 
already said in section 2.1.

In ref. \cite{jamin} it was also established that for the $I=3/2$ $K\pi$ S-wave 
scattering the unphysical cut contributions are not so small and one has to take 
care of them from the beginning. This is one of the reason why for this channel a 
description of the data `only' up to around 1.3 GeV could be given. This situation can 
also be applied to the $I=2$ S-wave $\pi\pi$ scattering. Detailed discussions about 
the quality of the experimental data are also given in ref. \cite{jamin} where some 
of the experimental ambiguities are discarded.  
 
\subsection{Final State Interactions}

As pointed out in ref. \cite{au} it is very interesting to complement the information 
coming directly from the strong interacting scattering  data with that obtained by the 
study 
of the Final State Interactions (FSI)
 due to the strong interactions between 
the produced mesons. 

It is then an important consistency check of the proposed solution to the S-wave puzzle 
to be 
able to reproduce in a systematic and unified way as many reactions as possible by 
taking care of the FSI. Along the years this program has been accomplished 
to a large extend and many reactions 
have been already reproduced: 
\vskip 15pt

\begin{enumerate}
\item[i)] $\gamma \gamma \rightarrow \pi^0 \pi^0$, $\pi^+\pi^-$, $K^0 \bar {K}^0$, 
$K^+ K^-$, $\pi^0 \eta$ ref. \cite{oo98}.
\vskip 10pt

\item[ii)] $\phi \rightarrow \gamma K^0 \bar{K}^0$ ref. \cite{plb} and $\phi \rightarrow 
\gamma \pi^0 \pi^0$, $\gamma \pi^0 \eta$ ref. \cite{toki}.
\vskip 10pt

\item[iii)] $J/\Psi \rightarrow \phi(\omega) \pi \pi$, $K\bar{K}$ ref. \cite{ulf2}.
\end{enumerate}  
\vskip 15pt

The $T$-matrix used to determine the FSI in the previous works is the one derived in 
ref. \cite{oo97} which is the limit case of ref. \cite{oo99} by keeping only 
the Adler zeroes. A detailed comparison between the $T$-matrices 
coming from refs. \cite{oo97,prd,oo99} is given in section III of the review 
\cite{review}. It is 
interesting to note that all these processes are related 
by unitarity and chiral symmetry and this is the reason why such a 
unified 
description of all of them has emerged (it is not just a matter of adding more and 
more uncorrelated
 free parameters as done in ref. \cite{au}). In fact, in many of the papers collected 
in i), ii) and iii), one uses the well known result that if a $T$-matrix is written as 
$N/D$ with $D$ having only 
the unitarity or right hand cut, which is absent in $N$, then a production mechanism 
without unphysical cuts can 
be written\footnote{Early considerations in these lines can be found in ref. 
\cite{gold}.} as $R/D$ with $R$ a function free of any cut, see also ref. \cite{ulf2}. 
The $R$ function is then fixed by requiring the matching with CHPT and/or making use 
of gauge invariance.  Note that FSI are tremendously 
important in all the above collected processes and can modify by orders 
of magnitude the Born term contributions.

\section{Chiral limit and chiral partners}

It is an interesting exercise to consider formally the limit $f_\pi \rightarrow 0$ in 
eq. (\ref{nolhc}). As it is 
well known $f_\pi \neq 0$ is a necessary and sufficient condition to have spontaneous 
chiral symmetry breaking. Hence, by studying the case $f_\pi\rightarrow 0$ we should 
obtain those 
features associated with the restoration of the chiral symmetry as for instance the 
appearance of particles degenerate in mass but with opposite parity. We will consider 
the SU(2) chiral limit, $m_u=m_d=0$ and $m_s$ fixed.

It is clear that as $f_\pi \rightarrow 0$ the chiral expansion no longer converges since 
the pion interactions become non-perturbative. However, to a large extend, this is 
precisely the situation actually observed in the $0^{++}$ $I=0,$ 1  and 1/2 channels  
for the physical value of $f_\pi$. We have handled this problem by introducing the
 non-perturbative scheme 
discussed along the manuscript. In the remaining of the section  we will assume that 
this scheme works as well in the limit $f_\pi \rightarrow 0$.


We have argued so far that a resummation of the CHPT series is very 
likely at the heart of the dynamical generation of the $\sigma$ meson, eq. 
(\ref{nolhc}). 
Specializing this equation to the case of only one CDD pole corresponding to the 
required Adler zero, see section 2.2, we have 
the following expression for the $I=0$ S-wave $\pi\pi$ amplitude:
\begin{equation}
\label{T}
T=\left[\frac{f_\pi^2}{s}+g(s) \right]^{-1}
\end{equation}

If we are interested in looking for a pole in $T$ we have to consider $g(s)$ in the 
second Riemann sheet, where it is given by:
\begin{equation}
g(s)=a^{SL}+\frac{1}{(4\pi)^2}\left(\log\frac{s}{\mu^2}+i\pi \right)
\end{equation}

A pole of $T$ is a zero of the denominator and hence one has the equation:

\begin{equation}
\label{sol}
\frac{f_\pi^2}{s_0}=-a^{SL}+\frac{1}{(4\pi)^2}\left(\log\frac{\mu^2}{s_0}-i\pi \right)
\end{equation}
with $s_0$ the corresponding solution.

If $f_\pi\rightarrow 0$ it is then convenient to write $s_0=\alpha f_\pi^2$ and 
therefore $\alpha$ must fulfill:

\begin{equation}
\label{higgs}
\frac{1}{\alpha}+\frac{\log \alpha}{(4\pi)^2}=-a^{SL}+\frac{1}{(4\pi)^2}\left( 
\log \frac{\mu^2}{f_\pi^2}-i\pi \right)
\end{equation}
Thus $\alpha=\alpha(\log\frac{\mu^2}{f_\pi^2})$ and the former equation admits the 
solution $\alpha(\log\frac{\mu^2}{f_\pi^2})\rightarrow 0$ when $f_\pi/\mu \rightarrow 0$.
\footnote{The same equation (\ref{sol}) is also found in the case of the strongly 
interacting Higgs sector when the massive states, that is, all the states except the 
vector gauge bosons, have a mass much larger than $4\pi f_\pi \approx 3$ TeV, here 
$f_\pi \equiv v\simeq 0.25$ TeV. In this case $\log\frac{\mu}{f_\pi}$ also goes to 
infinite because the scale $\mu$, which is related to the mass of the massive 
particles, is much larger than $f_\pi$. This scenario is discussed in ref. \cite{ww}.} 
Then $s_0=\alpha f_\pi^2$ vanishes in the limit $f_\pi \rightarrow 0$.

As a result, in the limit of chiral symmetry restoration one has:

\begin{table}
\begin{center}
\begin{tabular}{|l|c|c|}
\hline
State & $\pi$ & $\sigma$ \\
Mass  & $M_\pi=0$ & $m_\sigma=0$ \\
Parity & $-1$ & $+1$\\
\hline
\end{tabular}
\end{center}
\end{table}
and hence the dynamically generated $\sigma$ meson is the natural chiral partner 
of the pion.

The generalization of these considerations to the SU(3) case are straightforward and 
for instance in fig.9 of ref. \cite{oo99} the SU(3) set of dynamically generated poles 
are shown in the SU(3) chiral limit.

\section{Conclusions}

In this manuscript, we have demonstrated how the lowest 
lying scalar resonances appear as a consequence of chiral symmetry together with 
unitarity. 
It is also shown that crossed channel dynamics is suppressed in the SU(3) related 
S-waves with $I=0$, 1 and 1/2 in the pertinent energy region.

This picture has been tested already by post-dictions and also predictions of many 
production meson processes where FSI play a very important role giving rise to 
corrections of several orders of magnitude. 

Finally we have addressed the limit of chiral symmetry restoration (both spontaneous as 
well as explicit) and we have seen that the dynamically generated $\sigma$ meson is 
the required scalar degenerate in mass with the pions when chiral symmetry is restored.

\vskip 10pt

{\bf Acknowledgments}
\vskip 8pt
I am grateful to Ulf-G. Mei{\ss}ner for some pertinent remarks. I would like to 
thank as well E. Oset for a critical reading of the manuscript. This work has been 
supported in part by funds from DGICYT under contract PB96-0753 and from the EU TMR 
network Eurodaphne, contract no. ERBFMRX-CT98-0169.

\end{document}